\documentclass[twocolumn,tighten]{aastex63}

\usepackage{graphicx}
\usepackage{amsmath}
\usepackage{amssymb}
\usepackage{color}
\usepackage{mathtools}
\usepackage{ulem}
\usepackage[all]{hypcap} 

\usepackage{lineno}

\newcommand\msun{\, {M}_\odot}

\newcommand\gpcyr{\, {\rm Gpc}^{-3}\,{\rm yr}^{-1}}

\newcommand\msmbh{M_{\rm SMBH}}
\newcommand\mnsc{{M_{\rm NSC}}}
\newcommand\mimbh{M_{\rm IMBH}}
\newcommand\mgc{{M_{\rm GC}}}
\newcommand\fgc{{f_{\rm GC}^{\rm IMBH}}}
\newcommand\mgal{{M_{\rm *,gal}}}

\def\apjl{ApJL}%
\def\apjs{ApJS}%
\def\aap{A\&A}%
\begin{document}
\shorttitle{SMBH-IMBH mergers in NSCs}
\shortauthors{Fragione}

\title{Mergers of supermassive and intermediate-mass black holes in galactic nuclei from disruptions of star clusters}

\email{giacomo.fragione@northwestern.edu}

\author[0000-0002-7330-027X]{Giacomo Fragione}
\affil{Center for Interdisciplinary Exploration \& Research in Astrophysics (CIERA) and Department of Physics \& Astronomy, Northwestern University, Evanston, IL 60208, USA}
\affil{Department of Physics \& Astronomy, Northwestern University, Evanston, IL 60202, USA}

\begin{abstract}
Gravitational waves (GWs) offer an unprecedented opportunity to survey the sky and detect mergers of compact objects. While intermediate-mass black holes (IMBHs) have not been detected beyond any reasonable doubt with either dynamical or accretion signatures, the GW landscape appears very promising. Mergers of an IMBH with a supermassive black hole (SMBH) will be primary sources for the planned space-based mission LISA and could be observed up to the distant Universe. SMBH-IMBH binaries can be formed as a result of the migration and merger of stellar clusters at the center of galaxies, where an SMBH lurks. We build for the first time a semi-analytical framework to model this scenario, and find that the the comoving merger rate of SMBH-IMBH binaries is $\sim 10^{-4}\,\gpcyr$ in the local Universe for a unity IMBH occupation fraction, scales linearly with it, and has a peak at $z\approx 0.5$-$2$. Our model predicts $\sim 0.1$ event yr$^{-1}$ within redshift $z\approx 3.5$ if $10\%$ of the inspiralled star clusters hosted an IMBH, while $\sim 1$ events yr$^{-1}$ for a unity occupation fraction. More than $90\%$ of these systems will be detectable with LISA with a signal-to-noise ratio larger than $10$, promising to potentially find a family of IMBHs.
\end{abstract}

\section{Introduction}
\label{sect:intro}

The formation and evolution of the innermost galactic regions is still uncertain. Most of the observed galactic nuclei harbour supermassive black holes (SMBHs), with masses $\sim 10^5$ - $10^9$ \citep[e.g.,][]{FerrareseMerritt2000,korm2013}. Galaxies across the entire Hubble sequence also show the presence of nucleated central regions, the nuclear star clusters (NSCs). NSCs are generally very massive, with mass up to a few times $10^7\,\msun$, and very dense, with half-light radius of a few pc \citep[e.g.,][]{georg2016,NeumayerSeth2020}. In some galaxies, as our own Milky Way, SMBHs and NSCs are found to co-exist \citep[e.g.,][]{Capuzzo-DolcettaTostaeMelo2017}.

NSCs typically contain a predominant old stellar population, with age $\gtrsim 1$ Gyr, and show also the presence of a young stellar population, with age $\lesssim 100$ Myr \citep[e.g.,][]{BokervanderMarel2001,RossavanderMarel2006,CarsonBarth2015,MinnitiContrerasRamos2016,KacharovNeumayer2018}. While the latter requires some local recent star formation event, the former is comprised of stars as old as globular clusters (GCs). Therefore, a natural way to explain the origin of this population is through GC migration, and subsequent disruption, to the galactic center due to dynamical friction \citep[e.g.,][]{tremaine1975,capuzz2008,AntoniniCapuzzo-Dolcetta2012,Antonini2013,gne14}.

\begin{figure*} 
\centering
\includegraphics[scale=0.75]{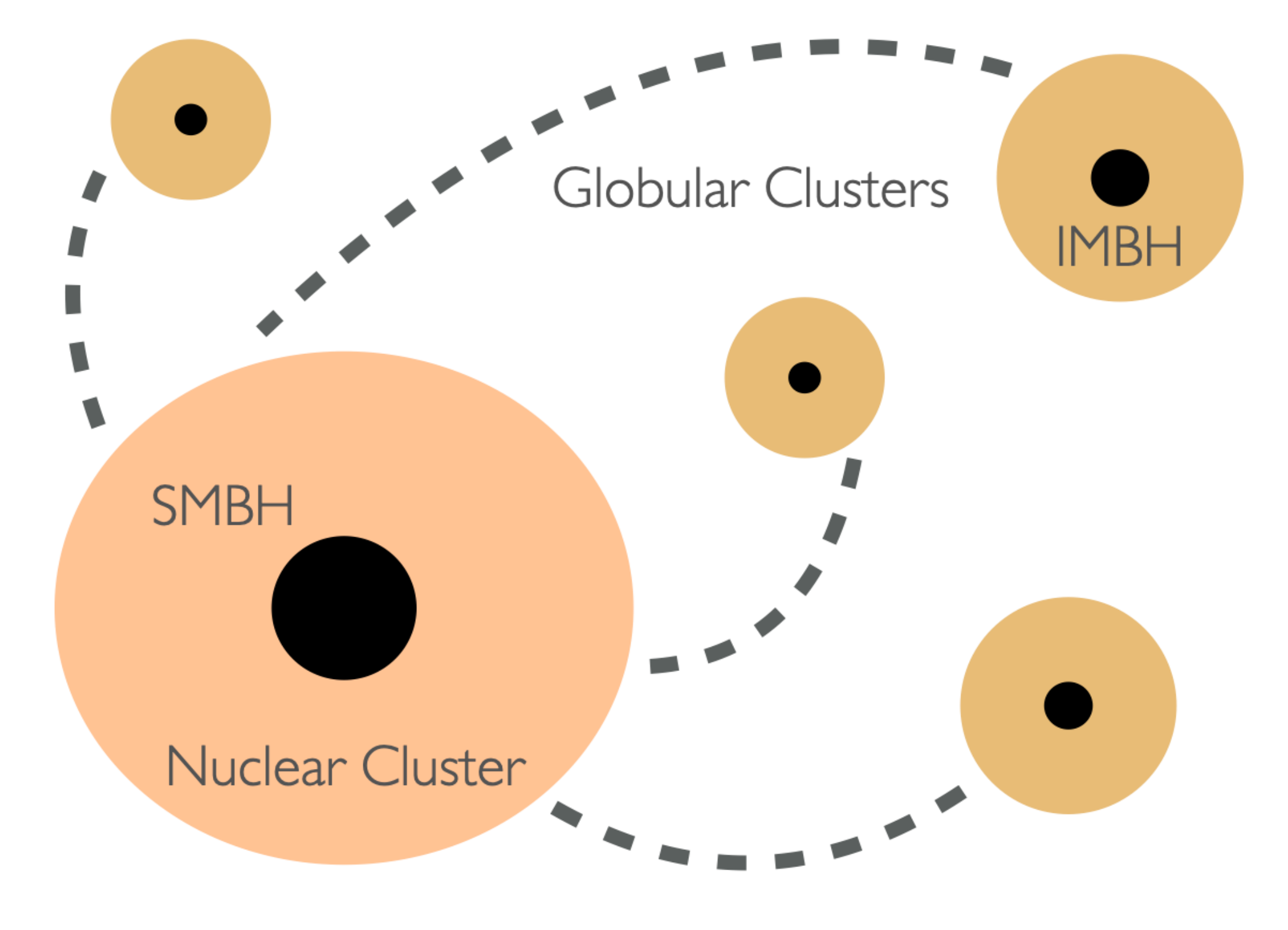}
\caption{Schematic illustration of the formation of an NSC as a result of GC migration and disruption due to dynamical friction. An SMBH sits in the center of the NSC, while IMBHs are assumed to be in the center of infalling GCs. As a result of cluster disruption, an SMBH-IMBH binary is formed, which will eventually merge via GW emission and be detectable with LISA.}
\label{fig:picture}
\end{figure*}

GCs represent a promising environment for forming intermediate-mass black holes (IMBHs), with masses in the range $\sim 10^2$ - $10^4\,\msun$. This would also be expected assuming that the observed relation between the SMBH mass and the velocity dispersion of stars around it holds at lower masses \citep[e.g.,][]{FerrareseMerritt2000,tremaine2002}. A number of studies has shown that a likely venue to form an IMBH is the so-called runaway scenario, in the early phases of cluster evolution. In this process, the most massive stars segregate and merge in the core of the cluster, forming a massive growing object that could later collapse to form an IMBH \citep[e.g.,][]{por02,gurk2004,frei2006,gie15,KremerSpera2020,DiCarloMapelli2021,GonzalezKremer2021}.

If an IMBH were to lurk in GCs that contribute to the assembly of NSCs, IMBHs would naturally be delivered to galactic nuclei in the proximity of an SMBH \citep[e.g.,][]{gurk2005,Mastrobuono-BattistiPerets2014,Arca-SeddaGualandris2018,FragioneGinsburg2018,fragleiginkoc18,Arca-SeddaCapuzzo-Dolcetta2019,AskarDavies2021}. The evolution of the SMBH-IMBH binary may depend on the specific orbit of the parent GC, on the details of the local stellar density profile, and on the number of IMBHs that are simulataneously delivered \citep{BaumgardtGualandris2006,PortegiesZwartBaumgardt2006,Mastrobuono-BattistiPerets2014,DosopoulouAntonini2017}. Eventually, the binary merges via gravitational wave (GW) emission (see Figure~\ref{fig:picture} for a schematic illustration).

SMBH-IMBH mergers will be primary sources for LISA and could be observed up to the distant Universe \citep[e.g.,][]{Amaro-SeoaneAudley2017,JaniShoemaker2020}. Despite their relevance, there have been only a handful attempts to model and compute the merger rate of SMBH-IMBH binaries resulting from migration and disruption of GCs in galactic nuclei \citep[][]{Arca-SeddaGualandris2018,Arca-SeddaCapuzzo-Dolcetta2019}. In this paper, we build for the first time a semi-analytical framework to model cluster disruptions and formation of SMBH-IMBH binaries, to compute their merger rates, and to assess their detectability with LISA. Our approach allows us to rapidly probe how the merger rates of SMBH-IMBH binaries are affected by galaxy masses, NSC and GC properties, and IMBH occupation fraction.

This paper is organized as follows. In Section~\ref{sect:method}, we discuss our numerical semi-analytical method to model SMBH-IMBH mergers. In Section~\ref{sect:results}, we present our results. Finally, in Section~\ref{sect:concl}, we discuss the implications of our findings and draw our conclusions.

\section{Method}
\label{sect:method}

In what follows, we describe the details of \textsc{simbhme}\footnote{\url{https://github.com/giacomofragione/simbhme}}, the numerical method we use to follow the formation and evolution of SMBH-IMBH binaries.

We start with sampling galaxy masses $M_{\rm *,gal}$ from a Schechter function
\begin{equation}
    \Phi(\mgal,z) = \Phi_*(z) \left(\frac{\mgal}{M_{\rm c}(z)} \right)^{\alpha_{\rm c}(z)} \exp\left(-\frac{\mgal}{M_{\rm c}(z)}\right)\,,
\label{eqn:galdist}
\end{equation}
from $M_{\rm *,gal}^{\rm min} = 10^{8.5}\msun$ to $M_{\rm *,gal}^{\rm max} = 10^{10.75}\msun$, corresponding approximately to the range where NSCs and SMBHs would co-exist \citep[e.g.,][]{Capuzzo-DolcettaTostaeMelo2017}. We set $\Phi_*(z=0)=0.8\times 10^{-3}$ Mpc$^{-3}$, $M_{\rm c}(z=0) = 10^{11.14}\msun$, and $\alpha_{\rm c}(z=0)=-1.43$, as extracted from the EAGLE cosmological simulations in \citet{FurlongBower2015}, which are consistent with the observed distribution of galaxies. While the galaxy distribution evolves as a function of redshift, we sample galaxy masses using the present-day distribution. This procedure ensures that the statistical distributions of galaxy masses, NSC masses, and SMBH masses in our model are consistent with the respective observed distributions in the local Universe. We take into account the redshift dependence of Eq.~\ref{eqn:galdist} when computing the merger rates (see Eq.~\ref{eqn:ratez}), where we reweigh the galaxy sample according to the galaxy distribution at a given redshift. For the evolution of the galaxy mass function as a function of redshift, see Table~A1 in \citet{FurlongBower2015}.

We use scaling relations from \citet{georg2016} for galaxies that host both an NSC and an SMBH, and compute the sum of their masses from
\begin{equation}
    \log((\mnsc+\msmbh)/\gamma_1)=\zeta\times\log(\mgal/\gamma_2)+\psi
    \label{eqn:mclmgal}\,,
\end{equation}
where $\gamma_1= 5.03\times 10^7\msun$, $\gamma_2= 2.76\times 10^{10}\msun$, $\zeta = 1.491$, $\psi=-0.019$. Note that this relation does not depend on the galaxy type. In sampling from Eq.~\ref{eqn:mclmgal}, we consider the scatter in the fit parameters. Then, to compute the SMBH  and NSC masses, we fit data in Figure~7 of \citet{georg2016} with
\begin{equation}
    \log(\msmbh/\mnsc)=A\times \log(\mgal)+B\,.
    \label{eqn:smbhnsc}
\end{equation}
From our least-square fit, we find $A=2.05$ and $B=-20.92$, with dispersion $\sigma=1.54$. Also in this case, we consider the scatter in the fit parameters when sampling from Eq.~\ref{eqn:smbhnsc}, and we discard from our analysis galaxies whose central SMBH or NSC would be less massive than $10^5\,\msun$.

We now compute the fraction of NSC mass formed as a consequence of GC migration and disruption, $f_{\rm out}$. We adopt the results of \citet{FahrionLeaman2021}, where a semi-analytical model of NSC formation based on
the orbital evolution of inspiraling GCs, together with observed NSC and GC system properties, was used to estimate the NSC mass formed \textit{in-situ} through local star formation. Following their approach, we first compute the mass formed \textit{in-situ}
\begin{equation}
    f_{\rm in} = \beta \tanh(\log(\mnsc)-\alpha)+(1-\beta)\,,
\end{equation}
where $\alpha=7.28$ and $\beta=0.34$ (dispersion $\sigma_{\rm in}=0.12$). Then, we simply estimate the fraction of mass accreted from GC disruptions as $f_{\rm out}=1-f_{\rm in}$.

Individual GC masses are sampled from the GC initial mass function, which we assume to be described by a negative power-law \citep[e.g.,]{Gieles2009,Larsen2009,ChandarWhitmore2010}
\begin{equation}
    f(\mgc)\propto M_{\rm GC}^{-2}\,,
\end{equation}
from $M_{\rm GC,min}=10^5\msun$ to $M_{\rm GC,max}=f_{\rm out}\mnsc$. We sample GC masses until the total sampled mass is $M_{\rm out}=f_{\rm out}\mnsc$, and draw cluster cosmic formation times from \citep[e.g.,][]{GrattonFusiPecci1997,GrattonBragaglia2003,VandenBergBrogaard2013,El-BadryQuataert2019}
\begin{equation}
    \psi(z)\propto \exp[-(z-z_{\rm GC})^2/\sigma_{\rm GC}]\,,
\end{equation}
where $z_{\rm}=3.2$ and $\sigma_{GC}=1.5$\footnote{We assume that the assembly of NSCs occurs within $\sim 10$-$100$\,Myr of the formation of GCs \citep[e.g.,][]{AntoniniCapuzzo-Dolcetta2012,Antonini2013,gne14}.}.

In dense star clusters, IMBHs could be originated mainly through repeated mergers either of massive main-sequence stars, later collapsing to form an IMBH, \citep{por02,gurk2004,frei2006,panloeb2012,gie15,TagawaHaiman2020,DiCarloMapelli2021}, or of stellar-mass BHs \citep{mil02b,OLeary+2006,antoras2016,antonini2019,FragioneKocsis2022,GonzalezKremer2021,MapelliDall'Amico2021,WeatherfordFragione2021}. Both processes depend on a number of initial cluster properties, including its density, primordial binary fraction, and the slope of the initial mass function. The mass distribution and the occupation fraction (that is the fraction of clusters that form an IMBH) of IMBHs is quite uncertain. For simplicity, we take the IMBH masses to be a fixed fraction
\begin{equation}
    \zeta = \frac{\mimbh}{\mgc}
\end{equation}
of the initial cluster mass. We consider different models with $\zeta=0.001$, $0.003$, $0.005$, and we take the IMBH occupation fraction in GCs, $\fgc$, to be $0.1$, $0.3$, $0.5$, $1.0$.

After the host GC delivers its central IMBH in the innermost regions of a galaxy, the formation, evolution, and eventual merger via GW emission of SMBH-IMBH binaries may depend on the specific orbit of the parent cluster, on the details of the local stellar density profile, and on the number of IMBHs that are simultaneously delivered \citep{BaumgardtGualandris2006,PortegiesZwartBaumgardt2006,Mastrobuono-BattistiPerets2014,DosopoulouAntonini2017}. For example, if the orbit of the parent cluster is not sufficiently elliptic, the emission of GWs could not be efficient in merging the SMBH-IMBH binary, or the interaction of IMBHs of $\sim 100\msun$ with the stars and compact objects surrounding the SMBH could quench the merger via GW emission \citep[e.g.,][]{ArcaSeddaBerczik2019,Arca-SeddaGualandris2018}. For simplicity, we assume that the delay time (from cluster formation to SMBH-IMBH merger) follows an exponential distribution with mean $\tau=1$\,Gyr \citep[][]{Arca-SeddaGualandris2018}. Note that this implies that the IMBH would have had enough time to be assembled between cluster formation and cluster disruption. This eventuality would be likely in the case the IMBH originates as a consequence of a rapid runaway process that could take place within $\sim 10$\ Myr from the cluster birth, while it would be more challenging if the IMBH grows primarily as a result of repeated mergers of stellar-mass BHs \citep[e.g.,][]{gie15,GonzalezKremer2021,FragioneKocsis2022}. To check how our results depend on the assumed distribution of delay times, we also run models with a $1/t$ distribution and a uniform distribution, with $t_{\min}=0.5$\,Gyr and $t_{\max}=10$\,Gyr being the minimum and maximum delay time, respectively.

\section{Results}
\label{sect:results}

\subsection{Mass and mass-ratio distributions}
\label{sect:massdist}

\begin{figure} 
\centering
\includegraphics[scale=0.585]{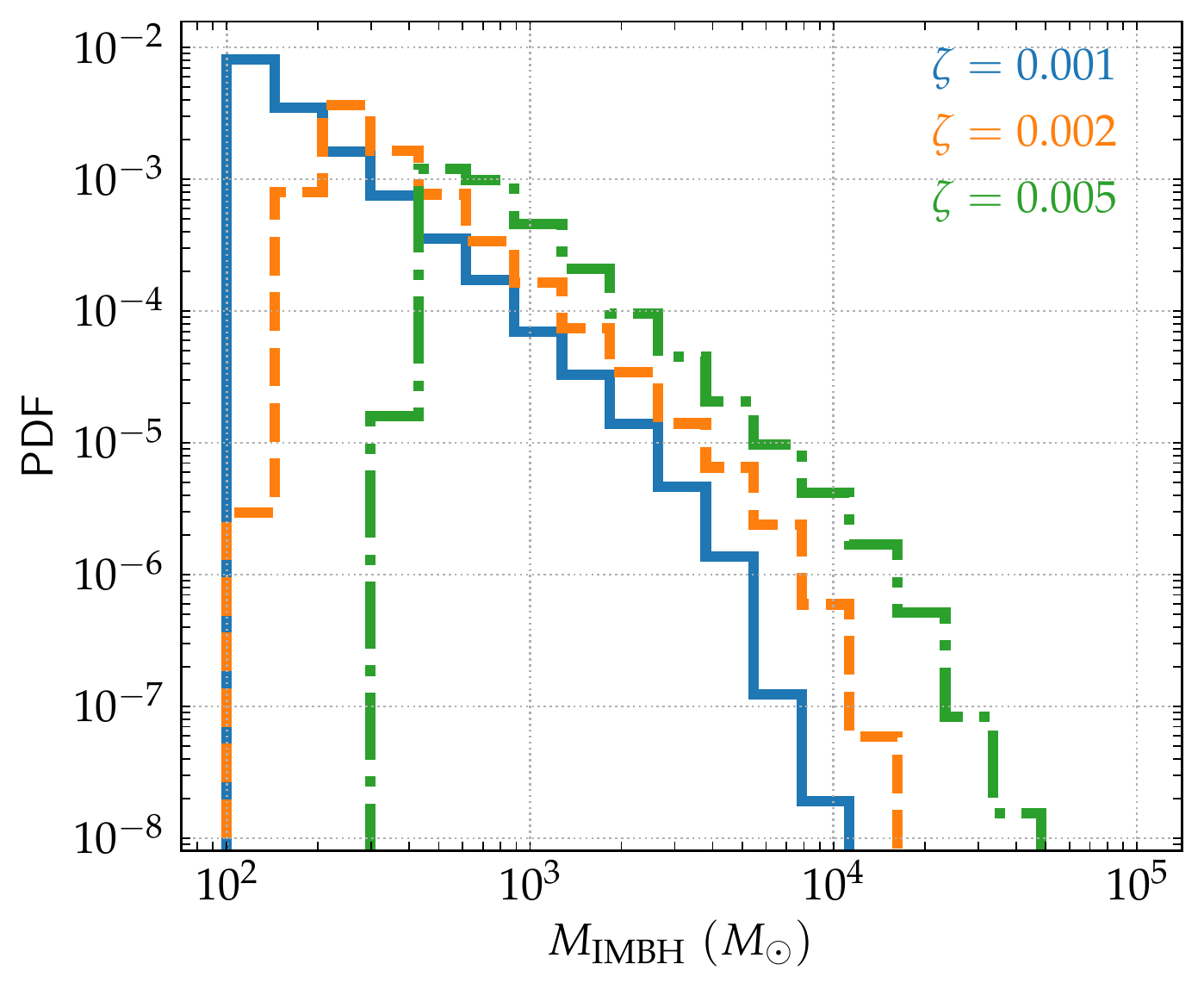}
\includegraphics[scale=0.585]{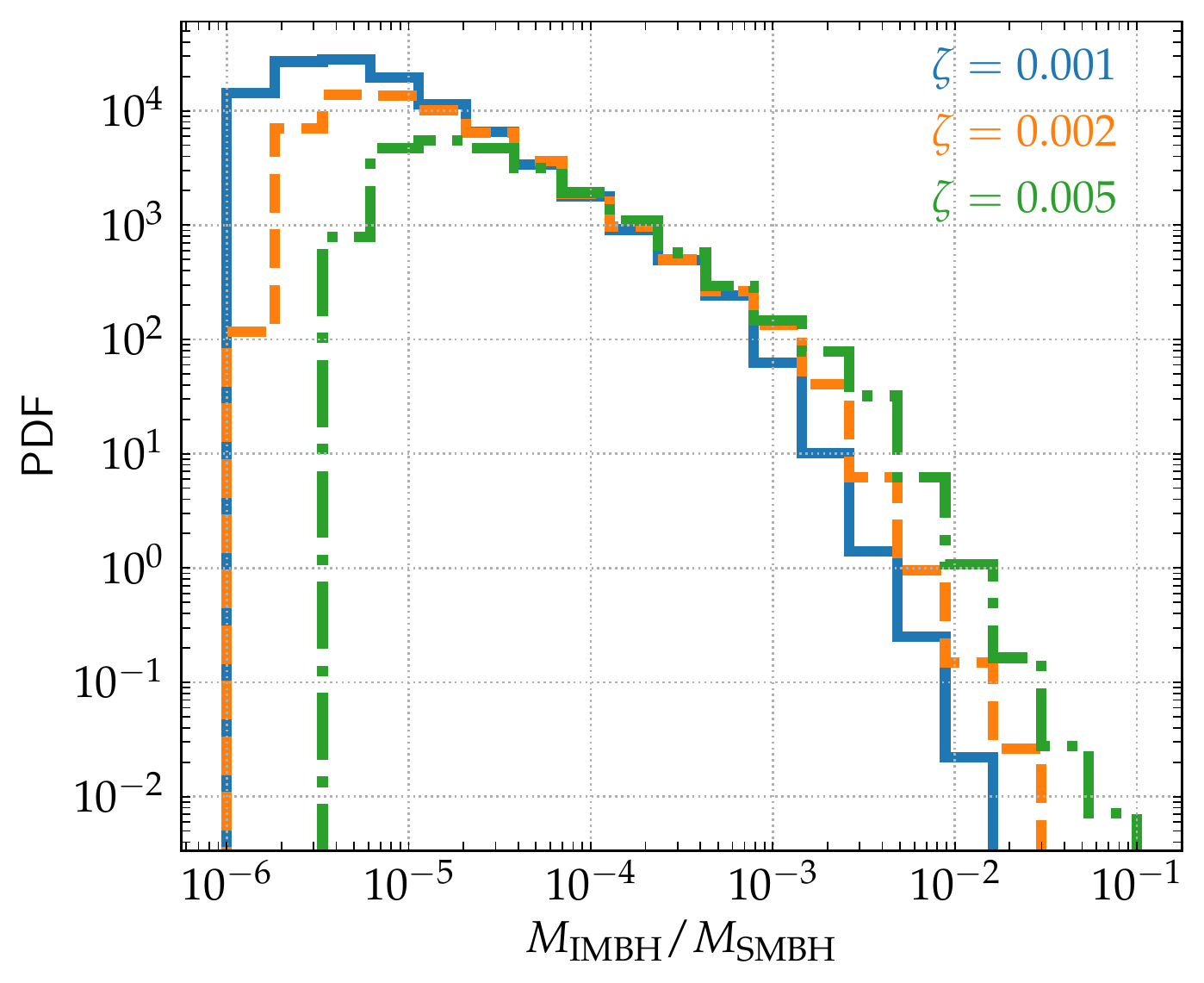}
\caption{Probability distribution functions of the mass of IMBHs (top) and mass-ratio of SMBH-IMBH binaries (bottom) that merge within a Hubble time. Different colors represent different assumptions on the IMBH mass, taken to be a fraction $\zeta$ of the mass of its parent cluster. The occupation fraction is fixed to $\fgc=1.0$ and the delay time follows an exponential distribution with mean $\tau=1$\,Gyr.}
\label{fig:masses}
\end{figure}

We show in Figure~\ref{fig:masses} the probability distribution functions of the mass of IMBHs in SMBH-IMBH binaries that merge within a Hubble time in our simulations for different assumption on the IMBH mass, taken to be a fraction $\zeta$ of its parent cluster mass. In these models, $\fgc=1.0$ and the delay time follows an exponential distribution with mean $\tau=1$\,Gyr. We find that about $50\%$ of the IMBHs that merge with an SMBH within a Hubble time have masses $\lesssim 200\,\msun$, $\lesssim 400\,\msun$, $\lesssim 1000\,\msun$ for $\zeta=0.001$, $\zeta=0.003$, $\zeta=0.005$, respectively. The slope of the distributions is independent of $\zeta$ and is $\propto M_{\rm IMBH}^{-2}$, as expected. Indeed, IMBH masses are scaled from GC masses, whose distribution is assumed to be $\propto M_{\rm GC}^{-2}$. 

The bottom panel of Figure~\ref{fig:masses} shows the probability distribution functions of the mass ratio $q=\mimbh/\msmbh$ of SMBH-IMBH binaries that merge within a Hubble time as a function of $\zeta$. We find that about $50\%$ of the binaries have mass ratios $\lesssim 7\times10^{-5}$, $\lesssim 1\times10^{-4}$, $3\times10^{-4}$ for $\zeta=0.001$, $\zeta=0.003$, $\zeta=0.005$, respectively.

\subsection{Merger rates}
\label{sect:rates}

We compute the merger rates as
\begin{equation}
\Gamma(z)= \epsilon \frac{d}{dt} \int_{M_{\rm *,gal}^{\rm min}}^{M_{\rm *,gal}^{\rm max}} \frac{\partial N(z)}{\partial M_{\rm *,gal}} \Phi(z,M_{\rm *,gal}) dM_{\rm *,gal}\,,
\label{eqn:ratez}
\end{equation}

where $\epsilon$ is the fraction of galaxies that host both an NSC and SMBH, and $\partial N(z)/\partial M_{\rm *,gal}$ is the number of mergers at a given redshift, $z$, per unit stellar galactic mass. The fraction of galaxies that host both an NSC and SMBH is highly uncertain and could depend on the galaxy mass and type \citep[e.g.,][]{NeumayerSeth2020}. In our calculations, we assume an average value of $\epsilon=0.3$\footnote{This value may evolve across cosmic time.}, from the semi-analytical models of \citet{AntoniniBarausse2015}.

\begin{figure} 
\centering
\includegraphics[scale=0.585]{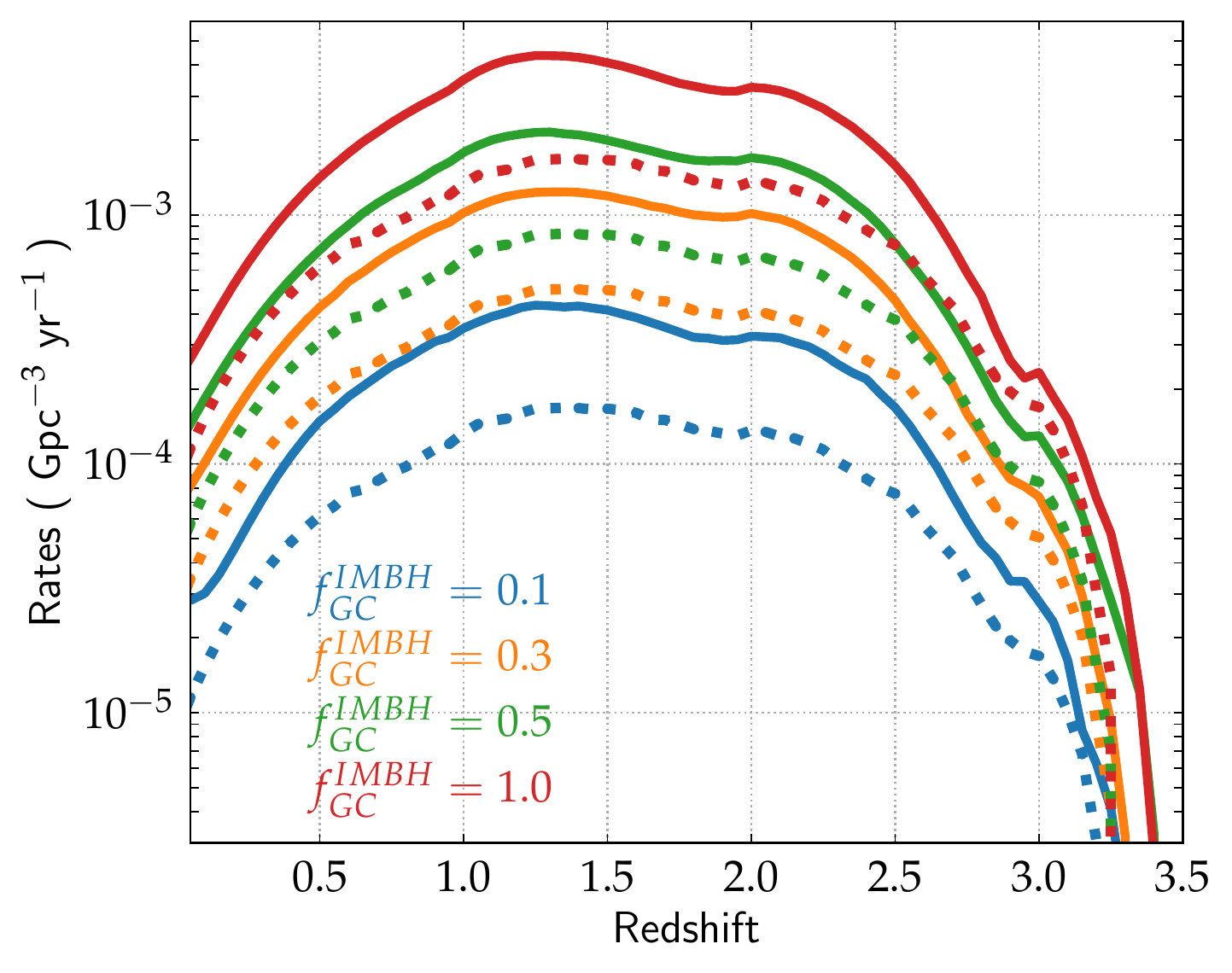}
\includegraphics[scale=0.585]{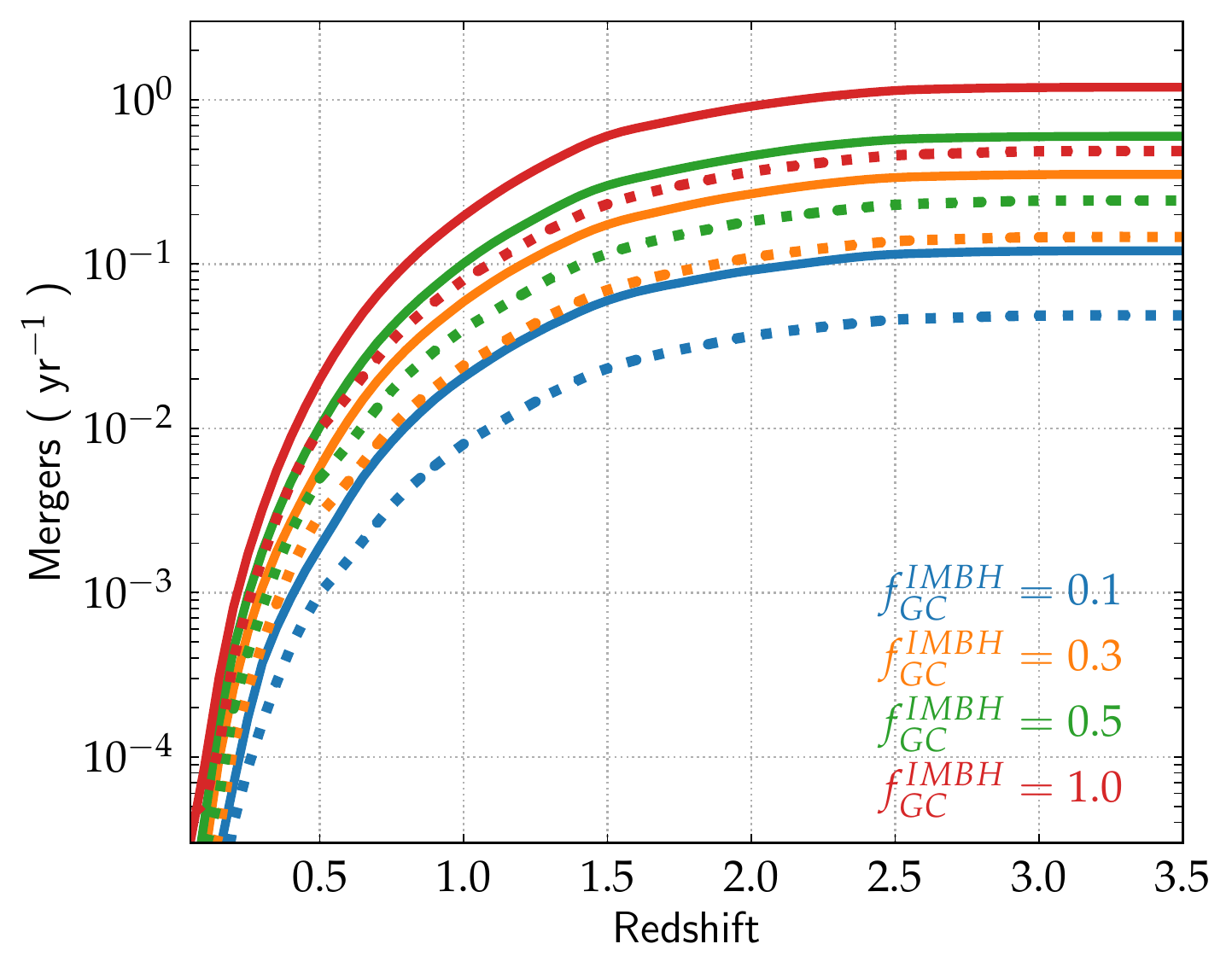}
\caption{Comoving (top) and cumulative (bottom) merger rates for SMBH-IMBH binaries from cluster disruptions in galactic nuclei. In these models, $\zeta=0.001$ and the delay time follows an exponential distribution with mean $\tau=1$\,Gyr. Different colors represent different IMBH occupation fractions. Solid lines: $M_{\rm GC,min}=10^5\,\msun$; dotted lines: $M_{\rm GC,min}$ from Eq.~\ref{eqn:mgcminscale} for late-type galaxies.}
\label{fig:rates_fgc}
\end{figure}

Figure~\ref{fig:rates_fgc} reports the comoving (top) and cumulative (bottom) merger rates for SMBH-IMBH binaries from cluster disruptions in galactic nuclei for different IMBH occupation fractions (solid lines). In these models, $\zeta=0.001$ and the delay time follows an exponential distribution with mean $\tau=1$\,Gyr. We find that the comoving rate is $\sim 10^{-4}\,\gpcyr$ in the local Universe for $f_{\rm GC}^{\rm IMBH}=1.0$, has a peak at $z\approx 2$, and scales linearly with the IMBH occupation fraction. When considering the cumulative rate, we find that our model predicts $\sim 0.1$ merger event yr$^{-1}$ within redshift $3.5$ if $10\%$ of star clusters harbour an IMBH, while $\sim 1$ yr$^{-1}$ if every cluster were to host an IMBH. Note that our results are consistent with the order-of-magnitude estimates in \citet{Arca-SeddaGualandris2018} and \citet{Arca-SeddaCapuzzo-Dolcetta2019}, who used a combination of semi-analytical estimates and $N$-body models of a limited sample of infalling star clusters.

We also compute merger rates for a different choice of the minimum GC mass. \citet{FahrionLeaman2021} argue that the minimum mass of GCs that inspiralled into the NSC scales $\propto R_{\rm eff}^2$, ($R_{\rm eff}$ is the effective radius of the galaxy) and can be approximately estimated by looking at the most massive cluster that has survived to present time. We adopt the scaling relation between a galaxy size and its mass from \citet{she03}
\begin{equation}
\log \left(\frac{R_{\rm eff}}{\mathrm{kpc}}\right)=\log b_1+ a_1\log \left(\frac{M_*}{M_\odot}\right)\ ,
\end{equation}
where $a_1=0.56$ and $b_1=3.47\times 10^{-5}$, for early-type galaxies, and 
\begin{eqnarray}
\log \left(\frac{R_{\rm eff}}{\mathrm{kpc}}\right)&=&\log c_2+a_2\log \left(\frac{M_*}{M_\odot}\right)\nonumber\\
&+&(b_2-a_2)\log \left(1+\frac{M_*}{M_0}\right)\ ,
\end{eqnarray}
where $a_2=0.14$, $b_2=0.39$, $c_2=0.1$, $M_0=3.98\times 10^{10}$ M$_{\odot}$, for late-type galaxies, and normalize to Milky Way's values, obtaining
\begin{equation}
M_{\rm GC,min}= 10^6\,\msun \left(\frac{R_{\rm eff}}{4\,{\rm kpc}}\right)^2\,.
\label{eqn:mgcminscale}
\end{equation}
We report our results in Figure~\ref{fig:rates_fgc} (dotted line) for late-type galaxies, and find that comoving and cumulative merger rates decrease by a factor of about $2$ with respect to case we assume $M_{\rm GC,min}=10^5\,\msun$. Instead, in the case we consider early-type galaxies, the rates decrease by a factor of about $4$.

\begin{figure} 
\centering
\includegraphics[scale=0.585]{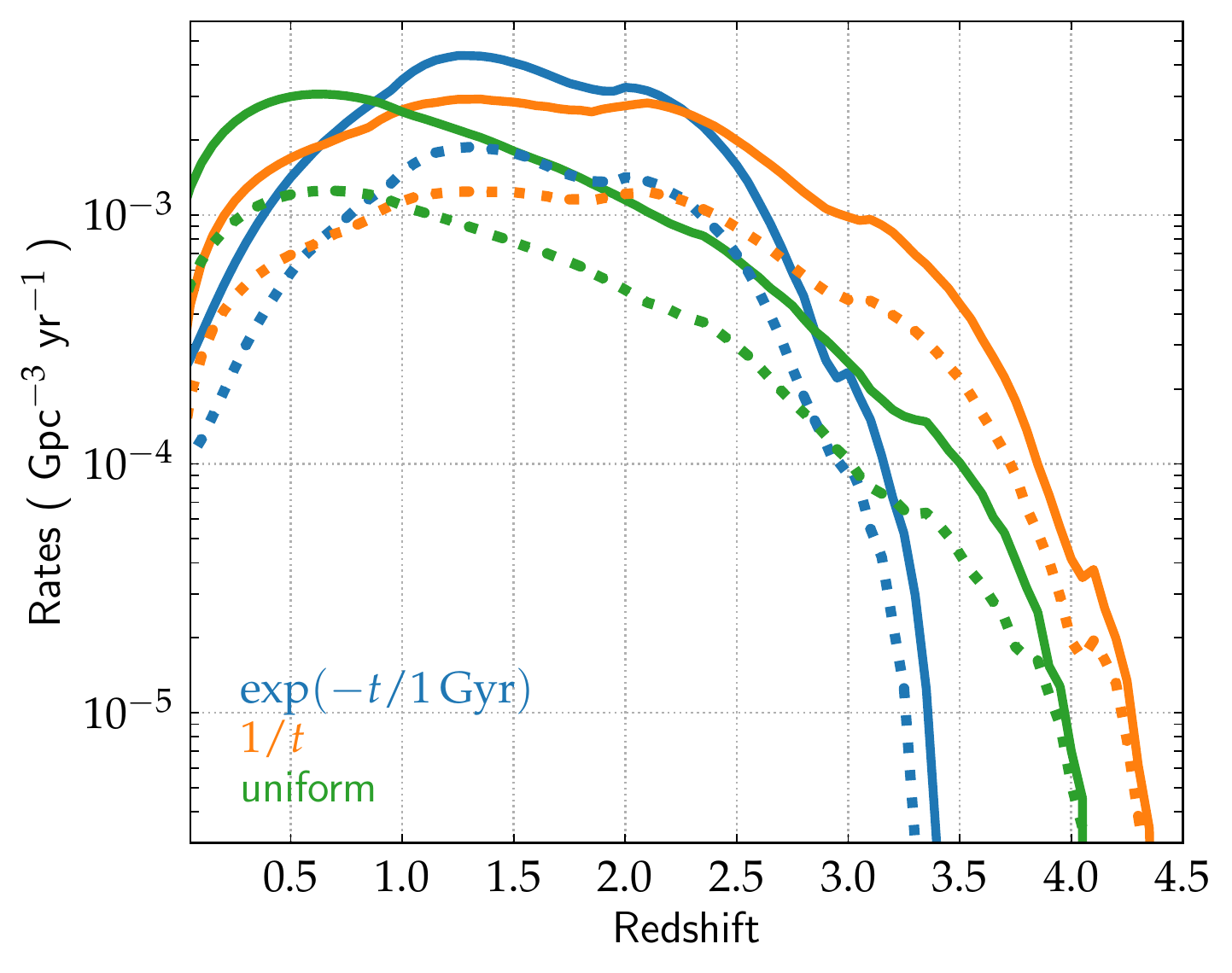}
\includegraphics[scale=0.585]{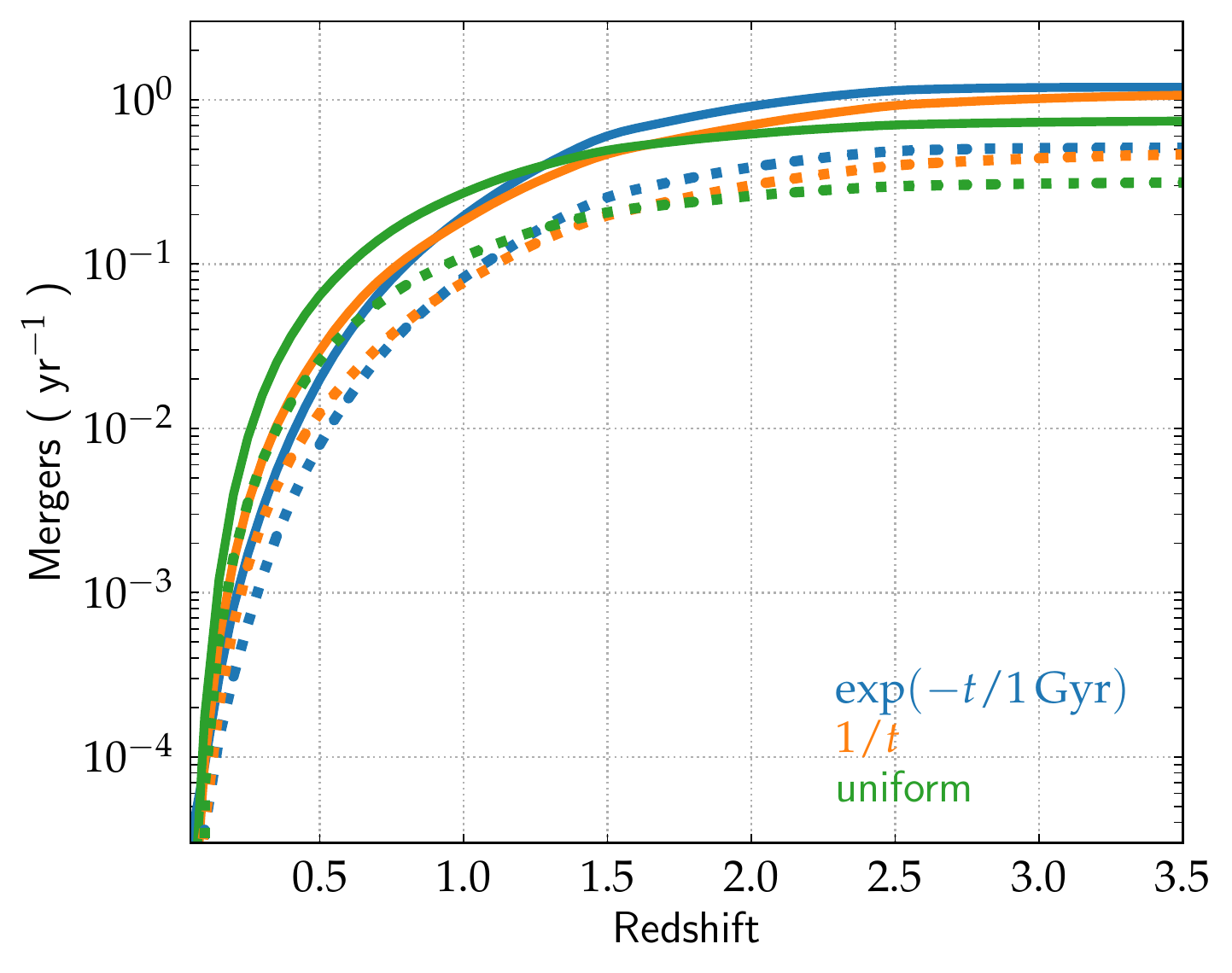}
\caption{Comoving (top) and cumulative (bottom) merger rates for SMBH-IMBH binaries from cluster disruptions in galactic nuclei ($\fgc=1.0$ and $\zeta=0.001$). Different colors represent different assumptions on delay times from cluster disruptions to SMBH-IMBH mergers. Solid lines: $M_{\rm GC,min}=10^5\,\msun$; dotted lines: $M_{\rm GC,min}$ from Eq.~\ref{eqn:mgcminscale} for late-type galaxies.}
\label{fig:rates_t}
\end{figure}

In Figure~\ref{fig:rates_t}, we report the comoving (top) and cumulative (bottom) merger rates for SMBH-IMBH binaries from cluster disruptions in galactic nuclei for different assumptions on delay times from cluster disruptions to SMBH-IMBH mergers both for $M_{\rm GC,min}=10^5\,\msun$ (solid lines) and $M_{\rm GC,min}$ from Eq.~\ref{eqn:mgcminscale} (dotted lines), assuming $\fgc=1.0$ and $\zeta=0.001$. Rates for different values of the occupation fraction can be easily obtained by considering that rates scales linearly with $\fgc$. We find that there is no significant difference in the total number of merger rates per year when we assume different distributions for delay times. However, while the comoving rate is peaked at $z\approx 2$ in the case of an exponential distribution, they are peaked at $z\approx 3$ for a $1/t$ distribution and at $z\approx 0.5$ for a uniform distribution, owing to typically shorter and longer delay times, respectively.

\subsection{LISA detections}
\label{sect:lisa}

We now compute the signal-to-noise ratio (SNR) for an SMBH-IMBH binary merger in LISA band. Note that these binaries enter the LISA band and merge within the nominal mission lifetime (of about $5$ yr).

We compute the average SNR as
\begin{equation}
\left\langle \frac{S}{N} \right\rangle=\frac{4}{\sqrt{5}} \sqrt{\int_{f_{\rm min}}^{f_{\rm max}} \frac{|\tilde{h}(f)|^2}{S_n(f)} df}\,,
\label{eqn:rhof}
\end{equation}
where $f_{\rm min}$ and $f_{\rm max}$ are the minimum and maximum frequency of the binary in the detector band, respectively, $S_n(f)$ is the effective noise power spectral density, and $|\tilde{h}(f)|$ is the frequency-domain waveform amplitude, approximated with a PhenomA waveform \citep[e.g., Eq.~20 in][]{robson2019}
\begin{eqnarray}
|\tilde{h}(f)| &=& \sqrt{\frac{5}{24\pi^{4/3}}} \frac{G^{5/6}}{c^{3/2}} \frac{M_{\rm c,z}^{5/6}}{D_{\rm L}f_0^{7/6}}\nonumber\\
&\times&
\begin{dcases}
(f/f_0)^{-7/6} & f<f_0\\
(f/f_0)^{-2/3} & f_0\leq f<f_1\\
w\mathcal{L}(f,f_1,f_2) & f_1\leq f<f_3\,,
\end{dcases}
\label{eqn:waveform}
\end{eqnarray}
where
\begin{eqnarray}
f_k &= & \frac{a_k\eta^2+b_k\eta+c_k}{\pi(GM_z/c^3)},\\
\mathcal{L} &=& \left(\frac{1}{2\pi}\right) \frac{f_2}{(f-f_1)^2+f_2^2/4},\\
w &= & \frac{\pi f_2}{2} \left(\frac{f_0}{f_1}\right)^{2/3}\,,
\end{eqnarray}
with $\eta=\msmbh\mimbh/(\msmbh+\mimbh)^2$, and the values of $\{f_k,a_k,b_k,c_k\}$ are reported in Table 2 in \citet{robson2019}. In Eq.~\ref{eqn:waveform}, $f_0$ is the observed (detector frame) frequency, related to the binary orbital frequency by $f_0=(1+z)^{-1}f_{\rm orb}$, $M_{\rm c,z}$ is the redshifted chirp mass, related to the rest-frame chirp mass
\begin{equation}
M_\mathrm{c}=\frac{\msmbh^{3/5}\mimbh^{3/5}}{(\msmbh+\mimbh)^{1/5}}
\end{equation}
by $M_\mathrm{c}=M_\mathrm{c,z}/(1+z)$, and
\begin{equation}
D_{\rm L}=(1+z)\frac{c}{H_0}\int_{0}^z \frac{d\zeta}{\sqrt{\Omega_{\rm M}(1+\zeta^3)+\Omega_\Lambda}}
\end{equation}
is the luminosity distance, where $z$ is the redshift, and $c$ and $H_0$ are the velocity of light and Hubble constant. We set $\Omega_{\rm M}=0.286$ and $\Omega_\Lambda=0.714$ \citep{planck2016}), respectively. We compute the power spectral density of LISA as in Eq.~1 in \citet{robson2019}.

\begin{figure} 
\centering
\includegraphics[scale=0.585]{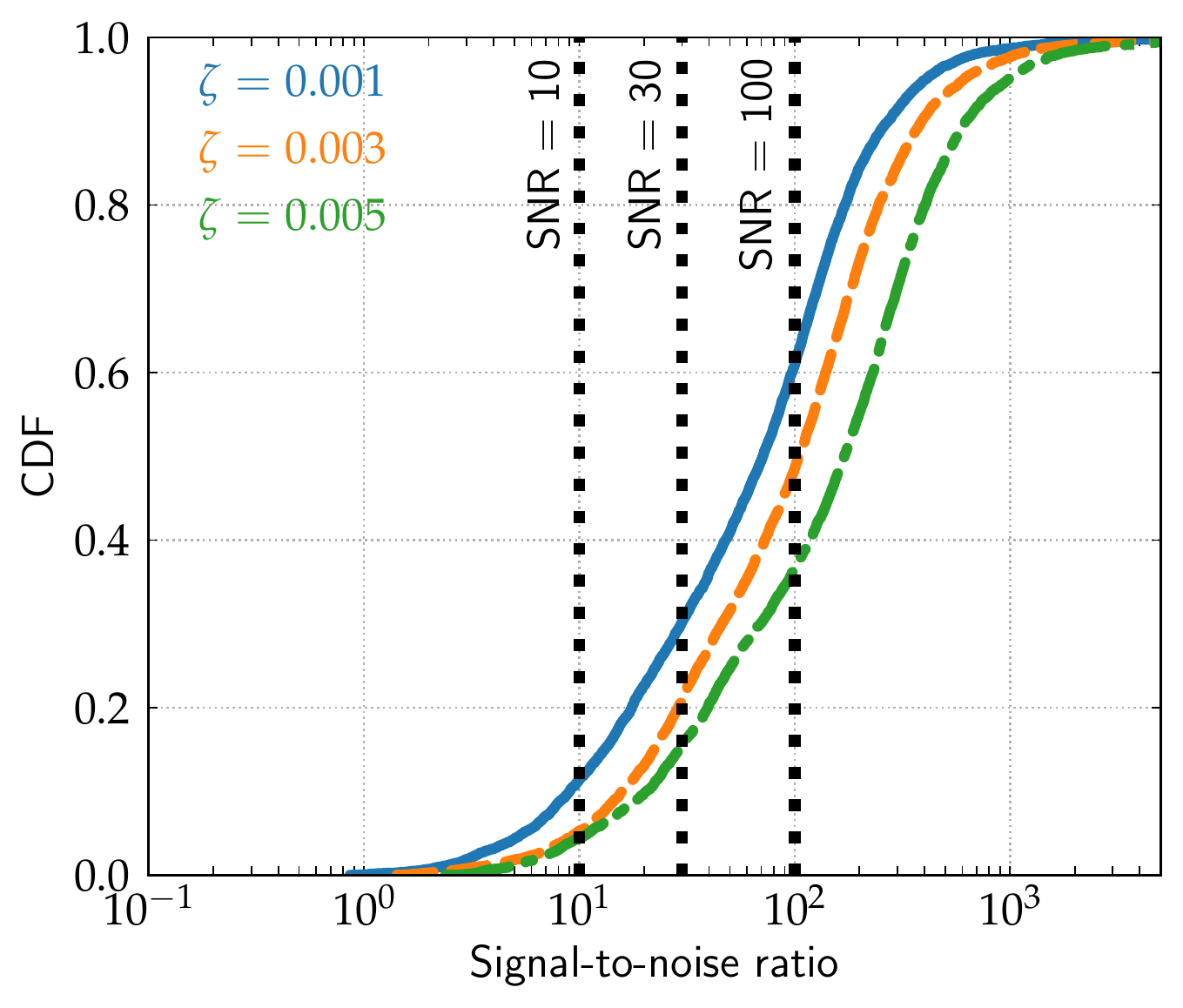}
\caption{Cumulative distribution function of the SNR in LISA for a cosmological population of SMBH-IMBH mergers from cluster disruptions in galactic nuclei. Different colors represent different assumptions on the IMBH mass, assumed to be a fraction $\zeta$ of its parent cluster.}
\label{fig:snr}
\end{figure}

We show in Figure~\ref{fig:snr} the cumulative distribution function of the SNR in LISA for a cosmological population of SMBH-IMBH mergers from cluster disruptions in galactic nuclei for different assumptions on the IMBH mass. In these models, $\fgc=1.0$ and the delay time follows an exponential distribution with mean $\tau=1$\,Gyr We find that about $90\%$, $80\%$, $50\%$ of the binaries have SNR larger that $10$, $30$, $100$, respectively, with a little dependence on $\zeta$. This clearly shows that LISA can detect with good confidence the large majority of the SMBH-IMBH mergers predicted in our model. 

\section{Discussion and Conclusions}
\label{sect:concl}

LISA offers a unique opportunity to discover IMBHs out to large redshifts and to make a big push forward in our understanding of their demographics. The implications of the possible existence of a large population of IMBHs in the Universe have only begun to be explored and only a handful of theoretical models that predict mass spectrum, redshift evolution, and merger rates of IMBH binaries have been developed.

We have built a semi-analytical framework to model cluster disruptions in galactic nuclei and formation of SMBH-IMBH binaries. We have shown that the comoving merger rate is $\sim 10^{-4}\,\gpcyr$ in the local Universe for a unity IMBH occupation fraction, scales linearly with it, and has a peak at $z\approx 0.5$-$3$, depending on the assumed distribution of delay times. Moreover, we have predicted $\sim 0.1$ event yr$^{-1}$ within redshift $z\approx 3.5$ if $10\%$ of star clusters host an IMBH, while $\sim 1$ yr$^{-1}$ for a unity occupation fraction. More than $90\%$ of these systems will be detectable with LISA with an SNR larger than $10$, with half of them being detectable with an SNR larger than $100$.

Our models represent an effective way, calibrated over the state-of-the-art results in the literature, to model the complex interplay between galaxy assembly, star cluster disruptions, and growth of the innermost regions of galaxies. Nevertheless, there are some limitations to our approach. In our models, we have assumed that the SMBH exists since the beginning of the NSC build-up. However, the mass of SMBHs evolves across cosmic time probably starting from smaller seed BHs, with the exact seeding mechanism not known. These seed BHs could have originated either from the evolution of massive stars or they may have formed with large initial masses through direct collapse of gas, and could subsequently grow from gas accretion, disruption of stars, and mergers with other BHs \citep[e.g.,][]{Volonteri2010,PacucciVolonteri2015,StoneKupper2017,FragioneSilk2020,Natarajan2021,AskarDavies2022}. Smaller SMBH masses would affect the distribution of stars and compact remnants, ultimately impacting the disruption of inspiralling star clusters and IMBH dynamics in the innermost region of the galaxy. Since the strain of the GW signal depends on the chirp mass of the merging binary (see Eq.~\ref{eqn:rhof}), smaller SMBH masses would also imply a smaller detectable volume for these type of events by LISA \citep{robson2019}. Another caveat of our approach is the lack of a detailed prescription for the evolution of any formed SMBH-IMBH, which may depend on the specific orbit of the parent cluster, on the details of the local stellar density profile, on the number of IMBHs that are simultaneously delivered, and the possible presence of gas \citep[e.g.,][]{BaumgardtGualandris2006,PortegiesZwartBaumgardt2006,Mastrobuono-BattistiPerets2014,DosopoulouAntonini2017}. Finally, the mass distribution and the occupation fraction of IMBHs in dense star clusters is quite uncertain, which depend on the cluster initial density, primordial binary fraction, and the slope of the initial mass function \citep[e.g.,][]{GonzalezKremer2021,WeatherfordFragione2021}. While the lack of confirmed IMBH detections in Galactic globular clusters seems to suggest a low occupation fraction \citep[e.g.,][]{GreeneStrader2020}, direct or indirect observations with next-generation observatories of the effect of IMBHs on the surrounding stars and compact objects will be crucial to constrain their numbers with high confidence \citep[e.g.,][]{GillTrenti2008,LeighLutzgendorf2014,BrightmanHarrison2016,PasquatoMiocchi2016,AnninosFragile2018,MezcuaCivano2018,BarrowsMezcua2019,WeatherfordChatterjee2020,WardGezari2021}. Despite the limitations of our framework, our model gives reasonable estimates of SMBH-IMBH mergers as a result of the inspiral of GCs, consistent with current order-of-magnitude estimates in the literature \citep{Arca-SeddaGualandris2018,Arca-SeddaCapuzzo-Dolcetta2019}.

A clear analysis of the role of all the above limitations and uncertainties could be given only by running a large set of $N$-body simulations, which are computationally expensive and beyond what current codes can handle. However, we can try to estimate and constrain the role of each of our assumption through the upcoming detections of GWs. Our models could be used to put an upper limit to the IMBH occupation fraction from the number of detected sources. For example, if LISA does not detect any SMBH-IMBH merger associated with galactic nuclei, our results would constrain the IMBH occupation fraction to $\fgc\lesssim 0.3$. With ever-enhanced sensitivity and with new detectors coming online, the characteristics of the IMBH family can be finally worked out. The forecast through the next decade includes tens or even hundreds of GW events, promising to shed light on the origin of this elusive population.

\section*{Acknowledgements}

We thanks the anonymous referee for useful comments. G.F.\ acknowledges support from NASA Grant 80NSSC21K1722. 

\bibliographystyle{aasjournal}
\bibliography{refs}

\end{document}